\documentclass[a4paper]{spie}  
\usepackage[]{graphicx}
\usepackage{units}
\usepackage{amsmath}
\usepackage{fancyvrb}
\newbox\verbbox
\title{PASSATA - Object oriented numerical simulation software for adaptive optics} 

\author{G. Agapito\supit{a}, A. Puglisi\supit{a}, S. Esposito\supit{a}
\skiplinehalf
\supit{a}Osservatorio Astrofisico di Arcetri, Largo E. Fermi 5, Firenze, Italy;
}

\authorinfo{Further author information: (Send correspondence to G. Agapito)\\G. Agapito: E-mail: agapito@arcetri.astro.it} 

\begin{document}

\def\arcsec{$^{\prime\prime}$}
\def\araa{Annual Review of Astronomy and Astrophysics}
\def\apj{The Astrophysical Journal}

\maketitle 

\begin{abstract}
We present the last version of the PyrAmid Simulator Software for Adaptive opTics Arcetri
(PASSATA), an IDL and CUDA based object oriented software developed in the Adaptive Optics
group of the Arcetri observatory for Monte-Carlo end-to-end adaptive optics simulations.
The original aim of this software was to evaluate the performance of a single conjugate adaptive optics system for ground based telescope with a pyramid wavefront sensor.
After some years of development, the current version of PASSATA is able to simulate several adaptive optics systems:
single conjugate, multi conjugate and ground layer,
with Shack Hartmann and Pyramid wavefront sensors.
It can simulate from 8m to 40m class telescopes, with diffraction limited and
resolved sources at finite or infinite distance from the pupil.
The main advantages of this software are the versatility given by the object oriented
approach and the speed given by the CUDA implementation of the most computational
demanding routines.
We describe the software with its last developments and present some examples of application.
\end{abstract}


\section{Introduction} \label{sec:intro}

PASSATA is an IDL based library/software capable of doing Monte-Carlo end-to-end Adaptive Optics (AO) simulations.
PASSATA was originally developed to evaluate the performance of the Large Binocular Telescope (LBT)
First Light Adaptive Optics (FLAO) system\cite{FLAO}.
This system is a single conjugate adaptive optics system with a pyramid wavefront sensor.
The first version was a library of functions called by a very long batch file with very limited
flexibility.
When the AO group of the Arcetri observatory started to work on other projects (for example: ARGOS\cite{rabienARGOS}, ERIS\cite{ERISSPIE2016}, GMT NGSAO\cite{doi:10.1117/12.2057059}),
PASSATA was rewritten to be more flexible and more user friendly. The library core was kept the same, but it was wrapped
in separated modules taking advantage of the possibilities of object-oriented programming made available by the
last versions of IDL. In addition, the most computationally demanding routines were reimplemented using GPUs.

\section{Structure} \label{sec:structure}

A PASSATA simulation is composed of any number of loosely coupled \emph{processing objects}, each of which is dedicated to a single task
(for example, apply a transformation from focal plane to pupil plane emulating an optical pyramid, or generating a PSF at a given wavelength based on an input wavefront).
Processing objects exchange data through \emph{data objects}, chosen from a list of pre-defined datatypes. Each processing object can define one or more data object as its output,
and has the responsibility of allocating it and filling its contents when new data has been generated. Most processing objects also need some kind of input data,
and any output data created by a processing object can act as an input to another processing object. It is the responsibility of the application programmer to connect
inputs to outputs using a standardized syntax to generate the proper processing chain.

It is not necessary for a processing object to produce data at every simulation step: typically, a processing object will check whether its input data has been refreshed,
and decides whether to process it immediately, or wait for more input at a later time. For example, a CCD processing object might integrate its input data for several simulation
steps, and produce output data (pixel frames) at a lower rate. Any processing object using the pixel data as an input will automatically slow down to follow the lower rate.
This makes it easy to simulate complex sensor setups where multiple channels are running at different rates.

In addition to processing and data object, a number of \emph{housekeeping objects} help in coordinating the simulation. Of these, the most important one is the \emph{loop control} object,
that keeps track of simulated time and triggers all processing objects once per simulation step. Other housekeeping objects assist in saving the simulation output, storing and
retrieving calibration data, and building the processing objects starting from a configuration file.

\subsection{Configuration} \label{sec:config}

All processing object parameters are stored in configuration files, called \emph{parameter files}.
Each parameter file is a collection of named IDL structures, with a slightly relaxed syntax (for example, there is no need of ending each and every line with the dollar sign).
The structure name is used to identify it throughout the code, while the content is used to initialize a specific processing object. Each structure field is evaluated with the EXECUTE function, and thus can contain any valid IDL code.

Although the library is capable of merging several parameter files, it is advised to keep all parameters for a simulation in a single master parameter file: when using
the standard data storage objects, this file is saved together with the rest of the simulation data, and allows to long-term tracking of the simulation parameters.

An interesting feature of the parameter file is the possibility of defining multiple values for one or more parameters, using a special syntax: when encountered, the library will generate multiple parameter files, assigning to each of them one of the possible parameter combinations. This way, it is extremely easy to start multiple simulations that explore several parameter ranges.

\section{Build and run a simulation loop}
\begin{figure}
\begin{center}
\includegraphics[width=12cm]{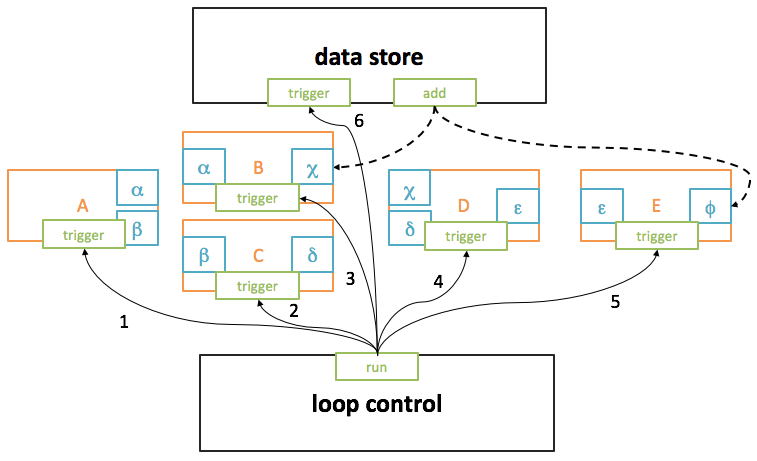}
\caption{Scheme of an example of simulation loop coordinated by loop control object.
		 Capital letters and orange color correspond to processing objects,
		 greek letters and cyan color correspond to data objects,
		 and green color correspond to object methods.
		 Data objects on the left of processing objects are inputs, while on the right are outputs.
		 Note that the can be output in one processing object and input in another one,
		 but they are not duplicated because only the reference to them is stored in processing objects.
		 The run order is shown by the numeration.}
\label{fig:loop}
\end{center}
\end{figure}
To build and run a simulation loop:
\begin{itemize}
\item write the parameters dictionary for each object
\item make a new factory object
\item call the factory methods to build all the processing and data objects needed from the parameters dictionary
\item set the input data objects of the processing objects
\item add in the correct order the processing objects to the loop control object
\item add to the data store the desired output objects
\item call the run method of the loop control object
\end{itemize}
When the run method is called a time step and a total time must be chosen.
The time step must be a submultiple of all the detectors integration time and a multiple
of the time resolution of the control loop object.\\
A scheme of a simulation loop is shown in Fig.\ref{fig:loop}.

\section{Example}

\begin{figure}
\begin{center}
\includegraphics[width=12cm]{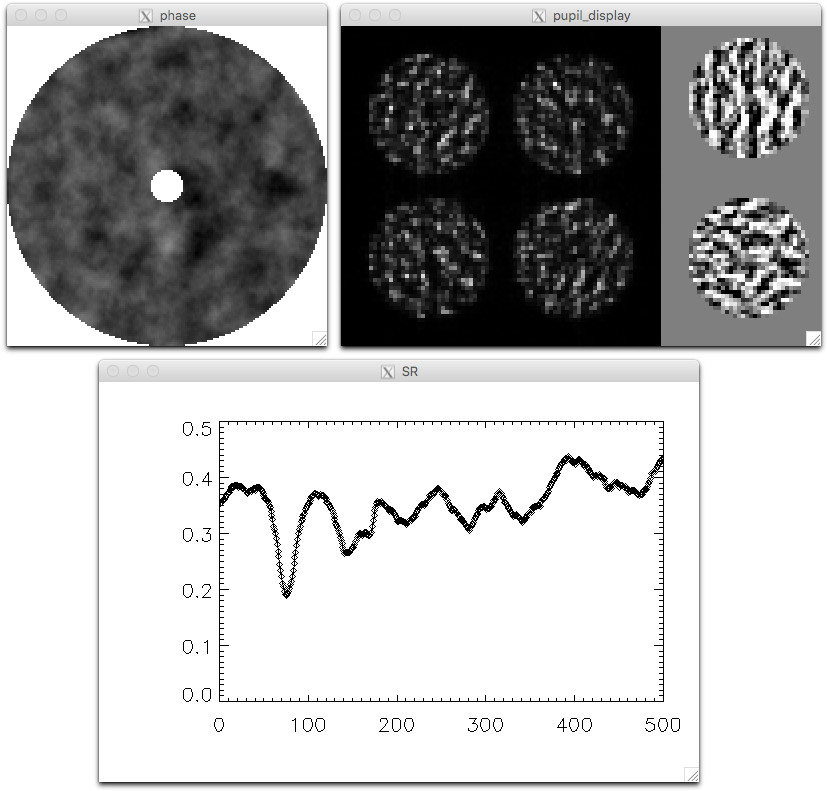}
\caption{Displays of the simulation example. From left to right and from top to bottom:
	residual phase on the Pyramid, detector frame and slopes and SR time history.}
\label{fig:simdisp}
\end{center}
\end{figure}

Here we describe how a generic simulation is built and run.
We choose as example a 8m-class telescope with a SCAO system.\\
This system comprises: a natural on axis source, an atmosphere made of 4 layers, a 30x30 sub-aperture pyramid WFS, a classic CCD detector, a real time computer -- which comprises a slope computer, a modal reconstructor and a modal integrator controller -- a DM capable to control 54 Zernike modes and a camera which compute H band PSFs.\\
The parameters and main file are shown respectively in table \ref{tab:param} and \ref{tab:main}.\\
An example of the displays of the simulation are shown in figure \ref{fig:simdisp}.\\
Note that a calibration of the pupils sub-apertures index, of the slope reference vector
and of the reconstruction matrix is required before running the simulation,
but only the code used to do the calibration of the pupils sub-apertures index
is reported in this work (see table \ref{tab:pupils}) for space reasons.
\setbox\verbbox=\vbox{\hsize=4in
{\scriptsize
\begin{verbatim}
{main,
 root_dir:          './SCAO',               ; Root directory for calibration manager
 pixel_pupil:       160,                    ; Linear dimension of pupil phase array
 pixel_pitch:       0.05,                   ; [m] Pitch of the pupil phase array
 total_time:        1.000d,                 ; [s] Total simulation running time
 time_step:         0.001d}                 ; [s] Simulation time step

{DM,
 type:              'zernike',              ; modes type
 nmodes:            54,                     ; number of modes
 npixels:           160,                    ; linear dimension of DM phase array
 obsratio:          0.1,                    ; obstruction dimension ratio w.r.t. diameter
 height:            0}                      ; DM height [m]

{pyramid,
 pup_diam:          30.,                    ; Pupil diameter in subaps.
 pup_dist:          36.,                    ; Separation between pupil centers in subaps.
 fov:               2.0,                    ; Requested field-of-view [arcsec]
 mod_amp:           3.0,                    ; Modulation radius (in lambda/D units)
 output_resolution: 80,                     ; Output sampling (corresponding to CCD pixels)
 wavelengthInNm:    750}                    ; [nm] Pyramid wavelength

{slopec,
 pupdata_tag :      'scao_pup',             ; tag of the pyramid WFS pupils
 sn_tag:            'scao_sn'}              ; tag of the slope reference vector

{control,
 delay:             2,                      ; Total temporal delay in time steps
 type:              'INT',                  ; type of control 
 int_gain:          1*replicate(0.5,54)}    ; Integrator gain (for 'INT' control)

{detector,
 size:              [80,80],                ; Detector size in pixels
 dt:                0.001d,                 ; [s] Detector integration time
 bandw:             300,                    ; [nm] Sensor bandwidth
 photon_noise:      1b,                     ; activate photon noise
 readout_noise:     1b,                     ; activate readout noise
 readout_level:     1.0,                    ; readout noise in [e-/pix/frame]
 quantum_eff:       0.32}                   ; quantum efficiency * total transmission

{wfs_source,
 polar_coordinate:  [0.0, 0.0],             ; [arcsec, degrees] source polar coordinates
 magnitude:         8,                      ; source magnitude
 wavelengthInNm:    750}                    ; [nm] wavelength

{camera,
 wavelengthInNm:    1650,                   ; [nm] Imaging wavelength
 nd:                8,                      ; padding coefficient for PSF computation
 start_time:        0.05d}                  ; PSF integration start time

atmo,
 L0:                40,                     ; [m] Outer scale
 heights:           [119.,837,3045,12780],  ; [m] layer heights at 0 zenith angle
 Cn2:               [0.70,0.06,0.14,0.10]}  ; Cn2 weights (total must be eq 1)

{seeing,
 constant:          0.8}                    ; ["] seeing value

{wind_speed,
 constant:          [5.,10.,20.,10.]}       ; [m/s] Wind speed value

{wind_direction,
 constant:          [90.,270.,270.,90.]}    ; [degrees] Wind direction value

{modalrec,
 recmat_tag:        'scao_recmat'}          ; reconstruction matrix tag

{pupil_stop,
 obs_diam:          0.1}                    ; pupil stop mask obstruction size     
\end{verbatim}
}
}
\begin{table}
\begin{center}
\begin{tabular}{|p{5in}|}
\hline
\box\verbbox
\\
\hline
\end{tabular}
\end{center}
\caption{SCAO parameters file.} \label{tab:param}
\end{table}
\setbox\verbbox=\vbox{\hsize=4in
{\scriptsize
\begin{verbatim}
; read parameters file
params = read_params_file('./params_scao.pro')

; Initialize housekeeping objects
factory = obj_new('factory',params.main, /GPU)
loop    = factory.get_loop_control()
store   = factory.get_datastore()

; Initialize processing objects
source  = list(factory.get_source(params.wfs_source))
atmo    = factory.get_atmo_container(source, params.atmo, $
              params.seeing, params.wind_speed, params.wind_direction)
prop    = factory.get_atmo_propagation(atmo, source)
pyr     = factory.get_modulated_pyramid(params.pyramid)
ccd     = factory.get_ccd(params.detector)
sc      = factory.get_pyr_slopec(params.slopec)
rec     = factory.get_modalrec(params.modalrec)
intc    = factory.get_control(params.control)
dm      = factory.get_dm(params.dm)
psf     = factory.get_psf(params.camera)

; Initialize display objects
sc_disp = factory.get_slopec_display(sc)                    ; slopes display
sr_disp = factory.get_plot_display(psf.out_sr)              ; SR display
ph_disp = factory.get_phase_display((prop.pupil_list)[0])   ; residual phase display
sc_disp.window = 10 & sr_disp.window = 11 & ph_disp.window = 12
sr_disp.title = 'SR' & sc_disp.disp_factor = 4 & ph_disp.disp_factor = 2

; Add atmospheric and DM layers to propagation object
atmo_layers  = atmo.layer_list
foreach layer,atmo_layers do prop.add_layer_to_layer_list, layer
prop.add_layer_to_layer_list, dm.out_layer

; Connect processing objects
pyr.in_ef           = (prop.pupil_list)[0]                  ; electric field
ccd.in_i            = pyr.out_i                             ; intensity objects
sc.in_pixels        = ccd.out_pixels                        ; pixel array objects
rec.in_slopes       = sc.out_slopes                         ; slopes objects
intc.in_delta_comm  = rec.out_modes                         ; modal residual objects
dm.in_command       = intc.out_comm                         ; modal commands objects
psf.in_ef           = pyr.in_ef                             ; electric field

; set data to be stored
store.add, psf.out_sr, name='sr'
store.add, pyr.in_ef, name='res_ef'

; Build loop
loop.add, atmo
loop.add, prop
loop.add, pyr
loop.add, ccd
loop.add, sc
loop.add, rec
loop.add, intc
loop.add, dm
loop.add, psf
loop.add, store
loop.add, sc_disp
loop.add, sr_disp
loop.add, ph_disp
 
; Run simulation loop
loop.run, run_time=params.main.total_time, dt=params.main.time_step

; add integrated PSF to store data
store.add, psf.out_int_psf.value

store.save, 'save_file.sav'
end
\end{verbatim}
}
}
\begin{table}
\begin{center}
\begin{tabular}{|p{5in}|}
\hline
\box\verbbox
\\
\hline
\end{tabular}
\end{center}
\caption{SCAO main file.} \label{tab:main}
\end{table}
\setbox\verbbox=\vbox{\hsize=4in
{\scriptsize
\begin{verbatim}
; read parameters file
dir = './'
params = read_params_file(dir+'params_scao.pro')

;increase modulation amplitude to improve the detection of the correct indices
params.pyramid.mod_amp = 20.

; Initialize housekeeping objects
factory       = obj_new('factory', params.main, /GPU)
loop          = factory.get_loop_control()
calib_manager = factory.get_calib_manager()

; Initialize processing objects
source  = factory.get_source(params.wfs_source)
prop    = factory.get_atmo_propagation(params.atmo, list(source))
pupstop = factory.get_pupilstop(params.pupil_stop)
pyr     = factory.get_modulated_pyramid(params.pyramid)
ccd     = factory.get_ccd(params.detector, params.pyramid)

; add pupil stop to propagation object
prop.add_layer_to_layer_list, pupstop

; Connect processing objects
pyr.in_ef = (prop.pupil_list)[0]
ccd.in_i  = pyr.out_i

; Build calibration loop
loop.add, prop
loop.add, pyr
loop.add, ccd

; Run calibration loop
loop.run, run_time=ccd.dt, dt=ccd.dt

; compute pupils from detector output
pupdata = pupil_acquire(ccd.out_pixels.pixels, thr1=params.slopec.thr1, $
                        thr2=params.slopec.thr2, no_show=~display)

; write pupils data on disk
calib_manager.write_pupils, params.slopec.pupdata_tag, pupdata

end     
\end{verbatim}
}
}
\begin{table}
\begin{center}
\begin{tabular}{|p{5in}|}
\hline
\box\verbbox
\\
\hline
\end{tabular}
\end{center}
\caption{Calibration of the pupils sub-apertures index file.} \label{tab:pupils}
\end{table}
\section{History/Reference works}
PASSATA has been intensively used by AdOpt group of Arcetri Observatory from 2008.
As said before, PASSATA has been developed for simulating SCAO system, like the LBT FLAO\cite{2010SPIE.7736E.116Q}
 the GMT NGSAO systems\cite{doi:10.1117/12.2057059} and EELT MAORY SCAO \cite{MAORYSALT2015},
 but in the last years the Arcetri observatory have worked on other AO techniques: 
\begin{itemize}
\item GLAO systems, like the LBT ARGOS system\cite{rabienARGOS}.
\item LGS SCAO systems, like the VLT ERIS AO system\cite{ERISSPIE2016} 
\item MCAO systems, like the study of a visible MCAO system\cite{MCAOVISSPIE2016}.
\item Extended (non point-like) reference source studies for SCAO and MCAO systems
\end{itemize}
For these studies the flexibility of PASSATA, given by the object-oriented programming, and the GPU acceleration
has greatly reduced the time needed to configure the new systems and to run the simulations.
A summary list of the simulated systems in this conference is:
\begin{itemize}
\item LBT SOUL system \cite{SOULSPIE2016}
\item VLT ERIS AO system \cite{ERISSPIE2016}
\item Visible MCAO system \cite{MCAOVISSPIE2016}
\item LGS system for WFS sensing study \cite{LGSSPIE2016}
\end{itemize}


\section{GPU acceleration} 
Many of the processing objects are computationally demanding. For this reason, we decided to offer the possibility of accelerating the code using GPUs, that can
run matrix algebra and FFTs much faster than traditional CPUs. We selected the CUDA API developed by NVIDIA\cite{Nickolls:2008:SPP:1365490.1365500}, mainly because of the very short learning curve. As a result, GPU acceleration is only possible with NVIDIA graphics cards. The CUDA functions are wrapped in a DLM extension that can be loaded by IDL at runtime. Effort has been concentrated on the most time-consuming routines, and as a result a few main areas have received most of the work:

\begin{itemize}
\item Wavefront propagation through phasescreens
\item Pyramid and Shack Hartmann propagation (from wavefront to detector plane)
\item Influence function multiplication (from DM commands to applied wavefront correction)
\item PSF generation from residual wavefront
\end{itemize}

Most of these functions have been built to take advantage of multiple GPUs if available in the same system. This has been done in-house, since the work started before this was a standard CUDA feature.

In PASSATA, GPU acceleration is optional. Each accelerated processing object still makes available an identical IDL implementation, and it is even possible
to mix GPU and non-GPU code in the same simulation: the \emph{factory} housekeeping object can build either the GPU or non-GPU version of each object, and when
connected together, most objects are capable of reading data from the GPU for local processing, or of uploading it to the GPU for the next processing step.
However, this mixing is discouraged, since transfer of data from the computer memory to the GPU (and vice-versa) is a very expensive operation. Typically,
the user does not need to manage such details: once the accelerated DLM is loaded, the system automatically detects it and builds GPU-accelerated versions
of the processing objects whenever available. In order to minimize data transfers, these objects will allocate GPU-aware data objects, that keep the data
on the GPU and just provide a \emph{handle} (an opaque identifier) to the GPU-allocated data that later processing objects can understand. In case a processing
object receives an input GPU data object but does not have a GPU implementation, the data is retrieved automatically from the GPU and made available to IDL.
It is also possible that the hardware used to run the simulation does not support part of the GPU routines: for example, the influence function (IF) multiplication needs to keep in GPU memory the whole IF matrix, that for an EELT-like system can easily reach 4-5 GB size. Not all GPUs have this much available memory. For cases like these, the \emph{factory} object employs a fallback allocation method: if enough memory is available, a GPU memory block is allocated. Otherwise, ordinary system memory is used, and a CPU routine is called in place of the GPU one. All this is transparent to the user, who just receives a warning if the allocator had to fallback to system memory.

\section{Conclusions}

The PASSATA simulator is now a flexible and powerful tool, that can simulate most existing or in-development AO systems, and that can be easily extended with new features, thanks to the object-oriented approach and the very loose coupling between different parts of the code. GPU acceleration provides typical speedups of 50x to 100x w.r.t CPU code, and allows the study of the largest AO system currently in development. The current bottleneck for simulator performance is the speed and memory capacity of currently available GPUs, and special cases like EELT systems with non-point like sources stretch the simulation time to the practical limit. However GPU vendors have encourangingly robust develoment roadmaps for the near future, and thus we expect these special cases to benefit from future hardware releases.


\bibliography{biblio}
\bibliographystyle{spiebib}

\end{document}